\documentclass[preprint,showpacs,preprintnumbers,amsmath,amssymb]{revtex4}

\usepackage{graphicx}
\usepackage{bm}

\def\t{\tilde}
\newcommand{\E}[1]{Eq.~(\ref{#1})}
\newcommand{\F}[1]{Fig.~\ref{fig:#1}}

\begin{document}

\preprint{APS/123-QED}

\title{Chemical turbulence equivalent to Nikolavskii turbulence} 

\author{Dan Tanaka}%
 \email{dan@ton.scphys.kyoto-u.ac.jp}
\affiliation{%
Department of Physics, Graduate School of Sciences, 
Kyoto University, Kyoto 606-8502, Japan
}%
\date{\today}%

\begin{abstract}   
We find evidence that a certain class of reaction-diffusion systems 
can exhibit chemical turbulence equivalent to Nikolaevskii turbulence. 
The distinctive characteristic of this type of turbulence is that 
it results from the interaction of   
weakly stable long-wavelength modes and 
unstable short-wavelength modes. 
We indirectly study this class of reaction-diffusion systems 
by considering an extended complex Ginzburg-Landau (CGL) equation 
that was previously derived from this class of reaction-diffusion systems. 
First, we show numerically that the power spectrum of this CGL equation 
in a particular regime is qualitatively quite similar to 
that of the Nikolaevskii equation.  
Then, we demonstrate that the Nikolaevskii equation can in fact 
be obtained from this CGL equation through a phase reduction procedure 
applied in the neighborhood of a codimension-two Turing--Benjamin-Feir point. 
\end{abstract}


\maketitle

The onset of spatiotemporal chaos 
is an important subject in the study of dissipative systems 
\cite{K84, Cro, Boh, Man}.  
Several yeas ago, 
a new mechanism causing the onset of spatiotemporal chaos was 
discovered by Tribelsky {\em et al.} \cite{Tri-discover} for  
the Nikolaevskii equation, 
\begin{equation}
\partial_{t}u=-\partial_{r}^{2}[\epsilon-(1+\partial_{r}^{2})^{2}]u-(\partial
_{r}u)^{2} \label{1}. 
\end{equation}
This equation was originally proposed to describe the propagation of 
longitudinal seismic waves in viscoelastic media \cite{Nik}. 
Its uniform steady state $u=0$ 
is unstable with respect to finite-wavelength perturbations 
when the small parameter $\epsilon$ is positive. 
However, this instability does not lead to spatially periodic steady states, 
because the equation possesses a Goldstone mode, 
due to its invariance under transformations of the form 
$u \rightarrow u + const.$,  
and the corresponding weakly stable long-wavelength modes 
interact with the unstable short-wavelength modes. 
As a consequence, 
spatially periodic steady states do not occur, and instead  
spatiotemporal chaos occurs supercritically. 
This chaos is called `Nikolaevskii turbulence'. 
Its properties have been investigated by several authors 
\cite{Tri1, Kli, Tri2, Xi-Amp, Mat, Tor, Fuj-Amp}.

Although it is conjectured that 
spatiotemporal chaos exhibiting a similar onset 
appears in various systems, 
experimentally 
only one such phenomenon has been observed to this time, 
complex electrohydrodynamic convection 
(also called `soft-mode turbulence'), 
discovered in homeotropically aligned nematic liquidcrystals
by Kai {\em et al.} \cite{Kai, Ros, Tri-liq_cry, Nag, Tam}. 
Similar onset has also been studied numerically 
in systems exhibiting Rayleigh-B$\acute{{\rm e}}$nard convection 
under free-free boundary conditions by Xi {\em et al.} \cite{Xi-RB}, 
and the possibility for the existence of this type of turbulence 
in reaction-diffusion systems has been investigated 
by Fujisaka and Yamada \cite{Fuj} 
and independently by Tanaka and Kuramoto \cite{Dan}.

In this letter, 
we present further evidence for 
the ubiquity of the type of spatiotemporal chaos  
described above. 
We find evidence that the chaos exhibited in a particular regime by 
a complex Ginzburg-Landau equation with nonlocal coupling \cite{Dan},
called a {\em nonlocal CGL equation}, 
is equivalent to Nikolaevskii turbulence.
This suggests that 
a certain class of oscillatory reaction-diffusion systems 
can also exhibit this type of chaos, 
because the nonlocal CGL is a reduced form of 
this class of reaction-diffusion systems \cite{K95, Dan}. 
These reaction-diffusion systems are such that
the chemical component constituing local oscillators are 
only weakly diffusive, 
while there is an extra diffusive component 
introducing an effectively nonlocal coupling between the oscillators.   
Such a situation, 
in which the coupling between the local oscillators is mediated 
by a diffusive substance,  
has been observed 
in some systems studied experimentally: 
biological populations, such as cellular slime molds \cite{bio2} and 
oscillating yeast cells under glycolysis \cite{bio4}, 
catalytic CO oxidation on metal surfaces \cite{Kim}, 
and the Belousov-Zhabotinsky reaction exhibited by a system dispersed 
in a water-in-oil aerosol OT microemulsion \cite{Van}. 
Because these systems possess such coupling, it is conjectured that 
under certain circumstances they could exhibit spatiotemporal chaos 
in the same class as Nikolaevskii turbulence.  


Our starting point is the following nonlocal CGL equation 
with complex amplitude $A$: 
\begin{eqnarray}
\partial_{t}A&=&A-(1+ic_{2})|A|^2 A+(\delta_1+i\delta_2)\nabla^2 A 
\nonumber\\
&& +K(1+ic_1) \int \!\! d{\bm r}'
G({\bm r}-{\bm r}')[A({\bm r}')-A({\bm r})]. \label{2}
\end{eqnarray}
Here $G$ is a coupling function, and 
$c_1$, $c_2$, $\delta_1$, $\delta_2$ and $K$ are real parameters.
In a previous paper \cite{Dan},  
this equation was derived as a generic reduced form 
of the class of reaction-diffusion systems discussed above. 
A simple example that belongs to this class is 
the following set of equations, describing  
a hypothetical extended Brusselator:   
\begin{eqnarray}
\partial_t X &=&
a - (b+1) X + X^2 Y + D \nabla^2 X + k c_X S , \\
\partial_t Y &=&
b X - X^2 Y + D \nabla^2 Y + k c_Y S , \\
\tau \partial_t S &=& -a -S + \nabla^2 S + X \label{rdS}, 
\end{eqnarray}
where the field variables $X$ and $Y$ 
represent limit-cycle oscillators 
existing immediately above the Hopf bifurcation, 
and the additional chemical component $S$ 
introduces an effective nonlocal coupling between these oscillators.  
The strength and anisotropy of this coupling are represented 
by the parameters $k$ and $c_{X,Y}$ respectively. 
The Hopf bifurcation parameter $\mu$ is 
written in terms of the parameters $a$ and $b$ 
as $\mu \equiv (b-b_c)/b_c$, where $b_c=1+a^2$.  
The coefficient $D$ is the diffusion constant for $X$ and $Y$,  
and the quantity $\tau$ is the time constant of $S$. 
When these couplings are as weak as the oscillation, i.e.,  
when $\mu \sim O(D) \sim O(k)$, this system can be reduced to \E{2}. 

Equation (\ref{2}) is invariant under transformations of the form 
$A \rightarrow Ae^{ic}$, with real constant $c$, 
which implies that
a uniform mode is neutrally stable.
This is a result of the spontaneous breaking of time translational symmetry
corresponding to the Hopf bifurcation
in the original reaction-diffusion systems. 
Also, 
the uniform oscillating solution of \E{2} possesses
a Turing instability \cite{Tur}
in a certain parameter region, as shown in \F{disp_on_bc}, 
in contrast to 
the ordinary diffusive complex Ginzburg-Landau equation, 
which possesses only a Benjamin-Feir instability.
Thus, \E{2} in this parameter region 
is characterized by weakly stable long-wavelength modes and 
unstable short-wavelength modes, like \E{1}.  
Therefore, it has been hypothesized that \E{2} exhibits 
spatiotemporal chaos similar to Nikolaevskii turbulence \cite{Dan}. 
We confirm this hypothesis in the following. 

\begin{figure}
\resizebox{0.4\textwidth}{!}{
\includegraphics{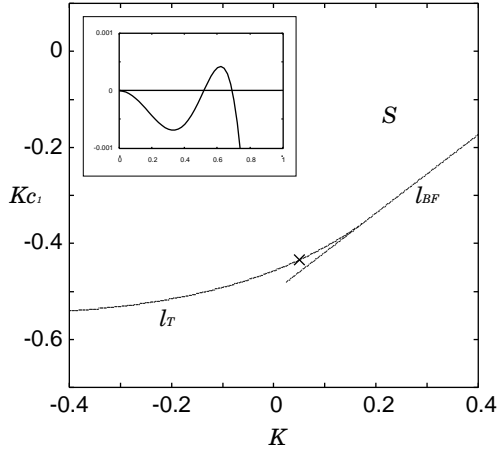}}
\caption{\label{fig:disp_on_bc}
Parameter space $(K, Kc_1)$ in the neighborhood of 
a codimension-two Turing--Benjamin-Feir point.
The uniform ocsillating solution of the nonlocal CGL is stable in domain $S$, 
possesses the Turing instablility above the curve $l_{T}$ and   
the Benjamin-Feir instablility above the line $l_{BF}$. 
The parameter values are $c2=1$, $\delta_1=0.3$, $\delta_2=0$, and 
$G(r)=\rho/2 \ e^{-\rho|r|}$, with $\rho=1+0.905i$. 
The inset plots  
the stability eigenvalue as a function of the perturbation wavenumber
for the uniform ocsillating solution 
at the point $(K, Kc_1)=(0.05, -0.4345)$, indicated by the cross. 
The largest eigenvalue is nearly equal to $0.0005$ 
at the Turing wavenumber, which is approximately $0.6$.
This dispersion curve is qualitatively the same 
as that of the Nikolaevskii equation. 
}
\end{figure}

For simplicity, we fix the parameter values as $c_2=1$ and $\delta_2=0$   
and consider only the case of one spatial dimension. 
In this case, the coupling function $G(r)$ is given by 
$G(r)=\rho/2 \ e^{-\rho|r|}$, with $\rho=1+i\eta$ and $1> \eta \geq 0$, 
where the form of $G(r)$ 
is derived from the original reaction-diffusion systems \cite{Dan}.
However, the following analysis is applicable also 
in higher-dimensional situations, in which $G$ takes other forms. 

A typical spatiotemporal pattern exhibited by \E{2}
in the slightly Turing-unstable regime 
close to the Benjamin-Feir criticality 
is shown in \F{shad}, 
where we see long-wavelength modulation of the Turing pattern.
The corresponding spatial power spectrum is displayed in \F{sk}.
This spectrum is found to have characteristic peaks 
at the Turing wavenumber and its harmonics.  
This feature is also seen in the spectrum of Nikolaevskii turbulence, 
shown in the inset of \F{sk}. 
This suggests that in the regime we consider, 
the spatiotemporal chaos exhibited by \E{2}  
is in the same class as Nikolaevskii turbulence. 
In the following, we show that, in fact, \E{1} can be obtained 
from \E{2} by means of a phase reduction technique \cite{K84}. 

\begin{figure}
\resizebox{0.35\textwidth}{!}{
\includegraphics{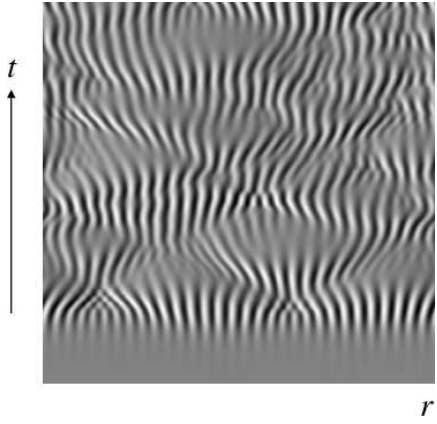}}
\caption{\label{fig:shad}
Gray level plots of $\Psi(r,t) \equiv \partial_r \arg[A(r,t)]$,  
which corresponds to $\partial_r u$ 
of the Nikolaevskii equation \E{1}.
The width of the longitudinal stripes corresponds to the Turing wavelength.
The parameter values are indicated by the cross 
in \F{disp_on_bc}. 
The configuration displayed here represents the result obtained after 
evolution from an initial configuration   
(which is the uniform ocsillating solution with a slight perturbation)     
over a time interval of length $10^5$ 
in a periodic one-dimensional system 
consisting of $2^{10}$ spatial points separated by a distance of $0.3$.
For the numerical scheme, we used 
a pseudo-spectral method with an 
explicit fourth-order Runge-Kutta scheme of time step $0.01$.
}
\end{figure}

\begin{figure}
\resizebox{0.35\textwidth}{!}{
\includegraphics{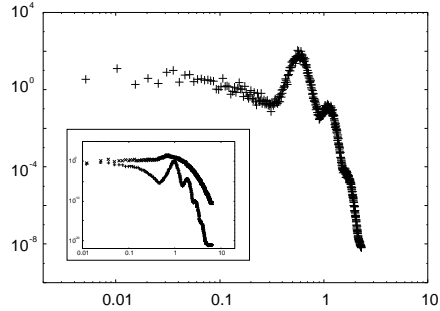}}
\caption{\label{fig:sk}
Spatial power spectrum of $\Psi(r,t)$. 
This spectrum was calculated using a system size 
($0.3 \times 2^{12}$ spatial points) larger than that of \F{shad}  
and averaged over a long time, 
excluding an initial transient of length $5 \times 10^4$. 
The inset displays typical spatial power spectra obtained from 
the Nikolaevskii equation, \E{1}, with $\epsilon=0.02$ (bottom) and 
the well-known Kuramoto-Sivashinsky equation (top),  
for the sake of comparison. 
The peaks are characteristic of Nikolaevskii turbulence. 
The spectrum of the nonlocal CGL exhibits similar peaks
at the Turing wavenumber and its harmonics. 
}
\end{figure}

In the phase reduction, 
the diffusion term and the nonlocal-coupling term of \E{2} 
are treated as perturbations. 
Using Floquet theory, 
we can calculate 
the operator ${\bm Z}(\phi)$ that projects the perturbation
onto the limit cycle $A^0(\phi)=e^{-i \phi}$ 
of the unperturbed system: 
${\bm Z}(\phi) = (\cos\phi-\sin\phi, -\sin\phi-\cos\phi)$.  
Hence, the phase dynamics of \E{2} obey 
\begin{eqnarray}
\partial_t \phi &=& 1 + \delta_1 \partial_r^2 \phi 
 - \delta_1 (\partial_r \phi)^2 
\nonumber\\&&
 + \int \!\! dr' \Gamma (\phi-\phi', r-r'), \label{6}\\
\Gamma(\phi, r)&=&
\frac{K|\alpha|}{2} e^{- |r|} 
[  \sin ( \phi -\eta \ |r|+\arg \alpha )
\nonumber\\ && \mbox{\hspace{1.5cm}}
  -\sin ( -\eta \ |r|+\arg \alpha )], 
\end{eqnarray}
where $\phi(r,t)$ and $\phi(r',t)$ are abbreviated as 
$\phi$ and $\phi'$, respectively, 
and $\alpha = [ -c_1-1 +i(1-c_1) ] \rho$.
In the integrand $\Gamma (\phi-\phi', r-r')$
there is a dependence on the quantity $\phi -(\phi'+\eta |r-r'|)$; i.e., 
the time evolution of the phase at $r$ depends not on the difference 
between the phases at $r$ and $r'$ 
but on the difference between the phase at $r$ and 
the phase at $r'$ plus the quantity $\eta |r-r'|$.
This might seem paradoxical for the following reason. 
If the spatial interactions are weak, then $\partial_t \phi \simeq 1$, and 
hence $\phi(r',t)+\eta |r-r'| \simeq \phi(r',t+t_0)$ with $t_0=\eta |r-r'|$.  
Thus the phase at $r$ interacts with the {\em future} phase $\phi(r',t+t_0)$, 
where $t_0 \geq 0$ because $\eta \geq 0$. 
To solve this paradox, note that 
the original reaction-diffusion fields oscillate roughly 
as $\cos\Phi$ with $\Phi = \omega_0 t - \mu \phi$, 
where the Hopf frequency $\omega_0$ is a finite positive value, 
and the Hopf bifurcation parameter $\mu$ is an extremely small positive value 
\cite{Dan}. 
Thus,
$\Phi(r,t)$ interacts with $\Phi(r',t-t_1)$ with $t_1=\mu t_0/\omega_0$ 
if we ignore terms of $O(\mu^2)$.  
Hence the field at $r$ interacts with 
the {\em past} field at $r'$, as $t_1 \geq 0$.
In this sense, 
$\eta$ producing the imaginary part in the coupling function $G$ in \E{2}
causes the phase coupling to be implicitly delayed, 
with the delay proportional to the distance between 
interacting oscillators.

Now we rescale the phase equation derived above.  
First, we introduce a variable $\psi$ defined as $\psi \equiv \phi-t$. 
The evolution equation for $\psi$ is the same as that of $\phi$, 
except for the absence of the term $1$ on the right-hand side. 
Instead of \E{6}, 
we refer to the equation for $\psi$ as the phase equation 
in the following. 
In the long-wavelength limit, 
where the validity of the phase description is ensured 
because the amplitude-like modes can be safely ignored, 
the phase equation can be expanded in a power series in $\partial_r^2$, 
owing to the symmetry of the system with respect to reflection through $r=0$: 
$\partial_t \psi = 
(\lambda_2 \partial_r^2 + \lambda_4 \partial_r^4 + \lambda_6 \partial_r^6 
+ \cdots )\psi + N$. 
Here $N$ represents the nonlinear terms, and $\lambda_n$ are constants 
derived from $c_1$, $\delta_1$, $K$ and $\eta$  
\footnote{
We have $\lambda_2 = \delta_1 + \sqrt{2} \Upsilon (\cos\zeta)^2 \cos(\vartheta -2\zeta-\pi/4)$ 
and 
$\lambda_{n \geq 4} = \sqrt{2} \Upsilon (\cos\zeta)^n \cos(\vartheta -n\zeta-\pi/4)$,  
with $\zeta \equiv \arctan \eta$ 
, where $\pi/4 > \zeta \geq 0$ because $1> \eta \geq 0$, and 
$K(1+ic_1) \equiv \Upsilon e^{i \vartheta}$,  
where $\Upsilon \geq 0$ and $2\pi> \vartheta \geq 0$.
}. 
We consider the Turing-unstable regime 
close to the Benjamin-Feir criticality, where 
$\lambda_2, \lambda_4, \lambda_6 > 0$, 
$\lambda_2 \sim O(\nu^2)$, $\lambda_4 \sim O(\nu)$, and $\lambda_6 \sim O(1)$, 
with the scaling parameter $\nu=+0$, 
and $4 \lambda_2 \lambda_6 = (1-\epsilon) \lambda_4^2$ 
with a positive constant $\epsilon \ll 1$. 
Under these conditions, 
the higher-derivative linear terms, $\partial_r^n \psi$ with $n \geq 8$,  
are much smaller than the other linear terms 
and can be ignored 
because the characteristic spatial scale of $\psi$ is $\sim O(\nu^{-1/2})$. 
Furthermore, because $\psi$ itself has 
a characteristic small magnitude depending on $\nu$, 
it is reasonable to assume that 
the largest nonlinear term in $N$ is 
$\gamma (\partial_r \psi)^2$, where $\gamma$ is a constant 
derived from $c_1$, $\delta_1$, $K$ and $\eta$  
\footnote{
$\gamma = 
[-\delta_1+\sqrt{2} \Upsilon (\cos\zeta)^2 \cos(\vartheta -2\zeta-3\pi/4)]$.  
}. 
In fact, 
when we rescale $r$, $t$ and $\psi$ as 
$r \rightarrow \t{r}=\sqrt{\frac{\lambda_4}{2\lambda_6}} r$, 
$t \rightarrow \t{t}=\frac{\lambda_4^3}{8\lambda_6^2} t$ and 
$\psi \rightarrow \t{\psi}=\frac{-4\gamma \lambda_6}{\lambda_4^2} \psi$, 
we can obtain the following scale-free equation from the phase equation: 
\begin{equation}
\partial_{\t{t}} \t{\psi} = 
-\partial_{\t{r}}^{2}[\epsilon-(1+\partial_{\t{r}}^{2})^{2}] \t{\psi} 
- (\partial_{\t{r}} \t{\psi})^2 \label{pNik}. 
\end{equation}
This is identical to \E{1}. 
Here, note that $\t{\psi}$ satisfies the scaling relation 
\begin{eqnarray}
\psi(r,t) &=&
 \nu^2 \t{\psi}(\nu^{1/2} r,\nu^3 t). 
\end{eqnarray}
This spatiotemporal scaling of the phase
is completely different from the scaling relation in the case that 
the Kuramoto-Sivashinsky equation is derived, 
$\psi(r,t) = \xi \t{\psi}(\xi^{1/2} r,\xi^2 t)$, 
where $\xi$ is a parameter that represents the Benjamin-Feir criticality. 

The above parameter conditions resulting in the derivation of \E{pNik} 
can be written in terms of the parameters of \E{2} as follows. 
Using $\zeta \equiv \arctan \eta$, 
where $\pi/4 > \zeta \geq 0$ because $1> \eta \geq 0$, 
\E{pNik} is derived from \E{2} under the following conditions, 
with a sufficiently small positive constant $\nu$  
when $\zeta \neq 0, \pi/8$:  
\begin{eqnarray}
K(1 + i c_1) &=& \Upsilon \exp[i(4\zeta+3\pi/4-\nu)] \label{Kc1}, 
\end{eqnarray}  
where
\begin{eqnarray}
\Upsilon&=&
\frac{\delta_1}{\sqrt{2} (\cos\zeta)^2}
\left[
\sin(2\zeta-\nu)+\frac{(1-\epsilon) (\sin\nu)^2}{4 \sin(2\zeta+\nu)}
\right]^{-1} . 
\end{eqnarray}
Here, the parameter $\epsilon$ is the same as that in \E{pNik}, and  
the critical value of $\Upsilon$ is 
$\Upsilon_c=\delta_1 [\sqrt{2} (\cos\zeta)^2 \sin(2\zeta)]^{-1}$.  

In order to observe 
the same spatiotemporal chaos as that of Nikolaevskii turbulence 
in our reaction-diffusion systems,  
we can analytically tune these parameters  
to satisfy the above conditions. 
In particular, 
the time constant $\tau$ of the additional chemical $S$
(e.g., see \E{rdS}) should be nonzero, 
because $\zeta=[\arctan(\omega_0 \tau)]/2 \neq 0$, 
where $\omega_0$ is the Hopf frequency \cite{Dan}.  
This reflects the importance of 
the imaginary part in the coupling function $G$ of \E{2}, i.e., 
the implicit delay of the phase coupling discussed above. 
This contrasts with the situation found in previous studies 
of similar reaction-diffusion systems, 
in which the limit $\tau \rightarrow 0$ was taken 
and $S$ was eliminated adiabatically 
\cite{K95, tau0-0, tau0-1, tau0-2, tau0-3}.  
In addition to the condition $\tau \neq 0$, we should also have 
$\tau \neq 1/\omega_0$, because $\zeta \neq \pi/8$.
This implies that the charcteristic time of the field $S$ is not comparable to 
that of the local oscillator, $(X, Y)$.  

In conclusion, 
we have found evidence that 
the nonlocal CGL and the corresponding class of reaction-diffusion systems
in a certain regime 
exhibit turbulence that is equivalent to Nikolaevskii turbulence. 
This result supports the conjecture of the ubiquitous nature of 
spatiotemporal chaos caused by 
the interaction between 
weakly stable long-wavelength modes and 
unstable short-wavelength modes.
We have confirmed through numerical calculation that these 
chaotic states are structurally stable for the nonlocal CGL
not just at the codimension-two Turing--Benjamin-Feir point 
but also in a finite neighborhood around it, at points where 
the CGL equation does not reduce exactly to the Nikolaevskii equation. 
Furthermore, 
we believe that similar spatiotemporal chaos would be found 
even if the phase-shift invariance were sligthly broken 
in the reaction-diffusion systems; i.e, 
this chaos should persist even if the Goldstone mode were lost. 
This conjecture is based on the observation that 
soft-mode turbulence exists even if one applies 
a small magnetic field that 
slightly breaks the arbitrariness of the azimuthal angle of directors 
bent through a Freedericksz transition 
in homeotropically aligned liquidcrystals \cite{Huh}. 
Finally, we note that the phase equation derived from the nonlocal CGL  
is a useful model 
for studying quantitative features of Nikolaevskii turbulence  
that has not yet been investigated sufficiently. 
Because this phase equation 
covers the transition region between 
Nikolaevskii turbulence and 
the well-known Kuramoto-Sivashinsky turbulence, 
it can be used not only for comparing these two types of turbulence  
but also for exploring the transition between them. 
This should lead 
to deeper understanding of Nikolaevskii turbulence 
and also a broader class of turbulence caused by 
interactions among modes with vastly different length scales.   

The author is grateful 
to Y.~Kuramoto for useful discussions,  
to S.~Kai, Y.~Hidaka, and K.~Tamura 
for valuable discussions on their experimental results, 
to H.~Fujisaka for interesting comments, and 
to H.~Nakao for carefully reading the manuscript.

\end{document}